\documentstyle[11pt,epsf]{l-aa}

\def\gs{\hbox{{\rm g s$^{-1}$}}}

\def\ergs{\hbox{\rm erg s$^{-1}$}}

\def\msun{\rm M_{\odot}}
\def\Teff{T_{\rm eff}}
\def\rtrans{r_{\rm tr}}

\def\apj{{ ApJ \/}}
\def\apjl{{ ApJ \/}}

\def\mnras{{ MNRAS \/}}
\def\acta{{Acta. Astron. \/}}
\def\aa{{ A\&A \/}}
\def\aass{{ A\&AS \/ }}

\def\pasj{{ PASJ \/}}

\def\al{{et al.\ }}
\newbox\grsign \setbox\grsign=\hbox{$>$}
\newdimen\grdimen \grdimen=\ht\grsign
\newbox\laxbox \newbox\gaxbox
\setbox\gaxbox=\hbox{\raise.5ex\hbox{$>$}\llap
     {\lower.5ex\hbox{$\sim$}}}\ht1=\grdimen\dp1=0pt
\setbox\laxbox=\hbox{\raise.5ex\hbox{$<$}\llap
     {\lower.5ex\hbox{$\sim$}}}\ht2=\grdimen\dp2=0pt
\def\gax{\mathrel{\copy\gaxbox}}
\def\lax{\mathrel{\copy\laxbox}}
\def\lta{\lax}
\def\gta{\gax}

\begin{document}

 \thesaurus{06         
              (02.01.2;  
               02.02.1;  
               02.09.1;  
               08.02.1;  
               13.25.5)  
               }
\title{Mechanisms for the outbursts of soft X--ray transients}

\author{J.-P. Lasota\inst{1,2}, 
R. Narayan\inst{3,4} and I. Yi\inst{5}}
      
       \offprints{J.P. Lasota (Meudon)}

\institute{
$^{1}${Belkin Visiting Professor, Department of Condensed Matter Physics,
Weizmann Institute of Science, 76100 Rehovot, Israel} \\
$^{2}${UPR 176 du CNRS; DARC, Observatoire de Paris, Section de 
Meudon, 92195 Meudon, France} \\
$^3${Harvard-Smithsonian Center for Astrophysics, 60 Garden Street,
Cambridge, MA 02138, USA} \\
$^{4}${Institute for Theoretical Physics, University of
California, Santa Barbara, CA 93106, USA}\\
$^{5}${Institute for Advanced Study, Olden Lane, Princeton,
NJ 08540, USA}
}
    \date{}

    \maketitle

\begin{abstract}
We show that the Keplerian thin disk in quiescent Soft X-ray
Transients cannot extend down to the last stable orbit around the
central black hole. We analyse the properties of the
Narayan, McClintock \& Yi (1996) model of quiescent Soft
X-ray Transients in which the cold Keplerian disk has its inner edge 
at a large transition radius and transforms to a hot, 
advection-dominated flow on the inside. We show that outbursts of
transient sources could be triggered in this model either by a pure
thermal accretion disk instability or by a disk instability generated
by an enhanced mass transfer from the stellar companion.  Both
mechanisms operate in the outer thin disk and could be at work, either
in different systems or in the same system at different epochs,
depending on the mass transfer rate and the value of the viscosity
parameter $\alpha_t$ of the thin disk.  We show that the recurrence
time between outbursts in SXTs can be explained with values of
$\alpha_t$ similar to these required by the dwarf nova disk instability
model instead of the unreasonably low values
needed in the model in which the thin disks extends down to the
last stable orbit. We extend the Narayan, McClintock \& Yi (1996) 
model to the case when the outer disk is non-stationary. We show that such
disk is too cold to account for the observed UV flux. This difficulty is
common to all models in which the outer disk is assumed to be optically
thick.


\end{abstract}

\section{Introduction}  

Soft X-ray Transients (SXTs) are close binary systems which undergo
large amplitude outbursts with a recurrence time of about $1 - 50$
years (see e.g. White 1994 for a review).  In most cases SXTs
disappear from the X-ray sky between outbursts leaving only the
companion star observable as a dim optical source.  Most SXTs are low
mass X-ray binaries (LMXBs), that is, binary systems in which a neutron
star or a black hole accretes matter lost by a Roche-lobe filling
low-mass stellar companion.  A minority of SXTs show X-ray bursts
during decline from outburst (e.g. Koyama et al. 1981) 
and must contain neutron stars.  In several
other SXTs, observations of the companion have allowed measurement of
a mass function which is larger than the maximum mass of a neutron
star (Orosz et al. 1994; Casares \& Charles 1994; Orosz et al. 1995;
Casares, Charles \& Marsh 1995; Bailyn et al. 1995).
These are certain to be black holes.  For the remaining SXTs there is
no direct evidence on whether they have neutron stars or black holes,
but it is believed that the majority of them contain accreting black
holes (see Tanaka \& Lewin 1995 and van Paradijs \& McClintock 1995).

Most of the outburst energy in an SXT is emitted in rather soft (1 to
few keV) X-rays. It has been claimed that the spectra of black--hole
SXTs (BSXTs) are characterised in outburst by an `ultra--soft'
(White et al. 1984)
component, and that this is not seen in neutron star SXTs.  However,
in a few cases involving good black hole candidates (e.g. V404 Cyg)
such a component has not been observed (Harmon et al. 1994). At least
one black hole SXT was observed to emit very high energy photons,
namely Nova Muscae 1991 which had an e$^{+}$--$e^{-}$ annihilation
line (Goldwurm et al. 1993). On the other hand some neutron star SXTs
have also been observed to emit above 100 keV (Barret et al. 1993).
It is therefore not clear if there are any spectral characteristics
that would allow, in general, to distinguish between neutron star and
black hole transients.

The binary periods of SXTs generally lie in the range between 5 hours
and 6 days.  (The `ultra-soft' transient system 4U 1630-47, [Parmar,
Angelini \& White 1995] could be an exception if its period is
confirmed to be 601.7$\pm 3.0$ days.)  SXTs are therefore LMXBs with
main--sequence dwarf companions or, at longer periods, (sub)giant
companions (King 1993). As in the case of their close cousins, the
dwarf novae, which have central white dwarfs instead of neutron stars
or black holes, two kinds of models have been proposed to explain the
outbursts of SXTs.

Hameury, King \& Lasota (1986) proposed a model in which outburts are due
entirely to a mass transfer instability in the X-ray illuminated
regions of the mass-losing companion. It was shown, however, by
Gontikakis \& Hameury (1993) that the existence of SXTs with orbital
periods under $\sim 10$ h is a problem since the model is unable to
produce the observed 
short ($\lta 10^6$ s) 
timescales (see also Lasota 1995).
 
The dwarf nova disk instability model (DIM)  
(see e.g. Cannizzo 1993) has been extended to the case of BSXTs by Huang \&
Wheeler (1989) and Mineshige \& Wheeler (1989).  Below we discuss
problems and difficulties that this model encounters when confronted
with observations.

In addition to the above two proposals, one should also consider
models in which the two mechanisms are at work in the same source.
For example, the disk instabilty could be triggered by an enhanced
mass transfer (EMT). Such ``mixed" or ``hybrid'' models have been
proposed for the superoutbursts of SU UMa type dwarf novae (Duschl \&
Livio 1989; Smak 1995,1996; Lasota, Hameury \& Hur\'e  1995) 
and suggested as a
possible explanation of SXT outbursts (Lasota 1996).

In this paper we analyse how various versions of the disk instabilty
model could be applied to outbursts of BSXTs. In section 2 we discuss
the difficulties of the dwarf nova DIM when it is applied to
BSXTs. Section 3 gives a short review of the advection-dominated flow
model of quiescent BSXTs (Narayan, McClintock \& Yi 1996). In section
4.1 we show that the `hybrid' EMT-triggered DIM could apply to
BSXTs. In section 4.2 we show that the outer cold disk in BSXTs could
be globally unstable and so BSXT outbursts could be triggered by a
thermal disk instability, as proposed independently by Mineshige
(1995).  We discuss problems and difficulties in section 5 and present
conclusions in section 6.

\section{The dwarf--nova type disk instability model for SXTs outbursts}

The DIM is based on the fact that accretion disk equilibria are
unstable at effective temperatures in the range $5000-10000$ K,
depending on the radial distance from the center.

It is conventional to discuss the instability in the $\Sigma-T_{\rm
eff}$ plane, where $\Sigma$ is the surface density of the disk and
$T_{\rm eff}$ is the effective temperature.  At a given radius, disk
equilibria trace a characteristic `$S$' shaped curve on this plane, as
indicated by the example shown in Figure~1 
(based on [Hameury, Hur\'e \& Lasota 1995]). 
The middle branch of the S
(shown by a dotted line) represents thermally and viscously unstable
equilibria and cannot represent physically accessible states.
Therefore, locally, an accretion flow has to choose between the two
stable branches of the S (solid lines), either the upper `hot' branch
or the lower `cold' branch. The upper branch exists only above a
minimum surface density $\Sigma_{\rm min}$ and minimum mass accretion
rate $\dot M_{\rm min} \propto
T^4_{\rm eff}$, where the value of $T_{\rm eff}$ corresponds to
$\Sigma_{\rm min}$, while the lower branch exists only below a certain
$\Sigma_{\rm max}$ and $\dot M_{\rm max}$.  Note that the limiting
surface densities and accretion rates vary with the radius $R$.

If the steady mass transfer rate $\dot M_T$ supplied by the secondary
star corresponds to conditions in the unstable middle branch, i.e. if
$\dot M_{\rm max}<\dot M_T<\dot M_{\rm min}$, then the disk cannot
maintain a steady state configuration and is forced into a local
`limit cycle' behaviour, alternating between the cold and hot
branches.  A dwarf nova outburst or SXT outburst would then correspond
to the transition from the cold to the hot state.  Observations
indicate that dwarf nova and SXT outbursts are global events where the
entire disk makes the transition.  In order to produce such a global
outburst, it is required that the pre-outburst quiescent disk be
entirely on the lower branch at all radii $R$, i.e. we must have
$\Sigma (R) < \Sigma_{\rm max} (R)$ at all $R$.

\begin{figure}
\epsfxsize=8.2 cm
\epsfbox[12 12 576 500]{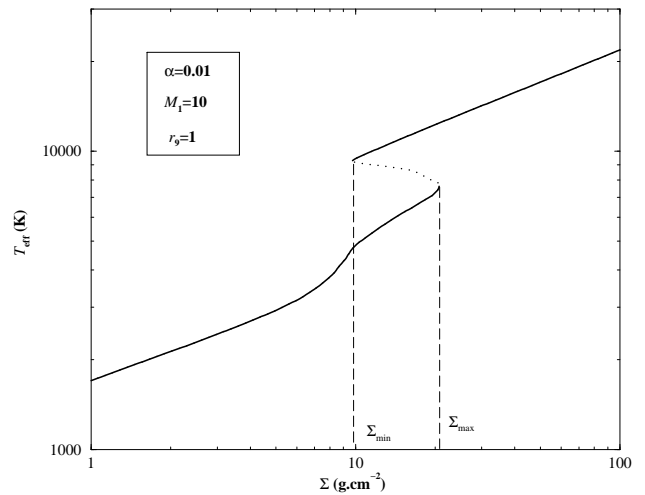}
\caption[]{Equilibrium solutions for a Keplerian accretion disk
at $R=10^{10}$ cm, for $M=10 \msun$ central mass and $\alpha_t = 0.01$.
The dotted part of the S-curve is thermally and viscously unstable.}
\end{figure}

In general, the critical density of the cold branch can be represented
as (Smak 1993):
\begin{equation}
\Sigma_{\rm max}= \Sigma_0 R^{b_r} M^{-b_r/3} \alpha_t^{b_{\alpha}},
\label{e2.1}
\end{equation}
where $M$ is the mass of the central object, $\alpha_t$ is the
Shakura-Sunyaev viscosity parameter of the thin disk, and the index
$b_r \approx 1$ so that $\Sigma_{\rm max}$ is an increasing function
of $R$. The $\alpha$ exponent is usually negative, $b_{\alpha} \lta
0$.

In a disk around a black hole in a binary system the radius may vary
by 5 to 6 orders of magnitude so that Eq. (\ref{e2.1}) would imply
very low densities in the quiescent inner disk.  For example,
introducing numerical values in Eq. (\ref{e2.1}) according to the
formula given by Smak (1993), we obtain
\begin{equation}
\Sigma_{\rm max}= 0.03 \ r^{1.11} m_{10}^{0.74} 
                   \left({\alpha_t \over 0.01}\right)^{-0.79} \ \
{\rm g \ cm^{-2}},
\label{e2.2}
\end{equation}
where $r = R/R_{\rm S}$ is the scaled radius in units of the
Schwarzschild radius $R_{\rm S}= 2GM/c^2$, and the mass is scaled
according to $ m_{10}= M/10\msun$.

According to the DIM, the quiescent disk in a dwarf nova or SXT system
is not in global equilibrium. Instead, the local accretion rate
increases with distance from the center (see Eq. (\ref{e2.5}) and the
effective temperature is constant with radius (see Eq. (\ref{e5.2}). This last
property is confirmed by observations (e.g. Wood et al. 1986; 1989).
Assuming that outbursts begin near the inner disk edge and taking
$\Teff = {\rm constant}$, Smak (1993) obtains for the recurrence time
\begin{equation}
                t_{\rm rec} \approx t_{\rm vis}= {2R_{\rm in}^2\over 63\nu},
\label{e2.3}
\end{equation}
where $R_{\rm in}$ is the radius of the inner edge of the disk and
$\nu$ is the kinematic viscosity coefficient.  In most BSXTs the
recurrence time is longer than $\sim 10$ yr.  Therefore, we obtain
from Eq. (\ref{e2.3}) a condition on the viscosity coefficient:
\begin{equation}
\nu \lta 276 \ t_9^{-1} r_{\rm in}^2 m_{10}^2 \ \ {\rm cm^2 \ s^{-1}} ,
\label{e2.4}
\end{equation}
where $t_9$ is the recurrence time in units of $10^9$ s.  Combining
Eq. (\ref{e2.2}) and Eq. (\ref{e2.4}), we then obtain the following
inequality which has to be satisfied by the accretion rate at the
inner edge of a quiescent disk if the DIM is correct (Lasota 1995):
\begin{eqnarray}
\dot M_{\rm in} &\lta & 3 \pi \nu \Sigma_{\rm max} 
\approx 
\nonumber \\  
& \approx & 83.5 \ t_9^{-1} r_{\rm in}^{3.11} m_{10}^{2.74} 
      \left({\alpha_t
    \over 0.01}\right)^{-0.79} \ \ {\rm \ g \ s^{-1}}.
\label{e2.5}
\end{eqnarray}
The DIM of Mineshige \& Wheeler (1989; MW) satisfies this inequality.

Eq. (\ref{e2.5}) provides an observational test of the DIM.  Several
BSXTs have been observed by GINGA (Mineshige \al 1992) and ROSAT (see
Verbunt 1995), and two systems, A0620-00 and V404 Cyg, have been
detected at levels corresponding to accretion rates (assuming an
efficiency of 0.1, see below) of at least $\sim 1.3 \times
10^{11}$~g~s$^{-1}$ for A0620-00 and $3 \times 10^{12}$~g~s$^{-1}$ for
V404 Cyg (Mineshige \al 1992; McClintock, Horne \& Remillard  1995; 
Verbunt 1995).
For reasonable values of $\alpha_t\sim10^{-2}$, these accretion rates
are several orders of magnitude larger than the values implied by
Eq. (\ref{e2.5}) for $r_{\rm in}=3$ (inner edge at the last stable
orbit as in the dwarf--nova type model).  Therefore, the detections of X-ray
radiation from quiescent SXTs contradict the disk instability model,
at least in the version proposed by MW (Mineshige \al 1992).

We note that in the case of A0620--00, the prototypical BSXT, the
optical luminosity suggests a mass accretion rate of $\sim 6 \times
10^{15}$~g~s$^{-1}$ in the outer part of the accretion disc
(McClintock et al. 1995), at $r>10^4$. 

\section{Advection-dominated flows in quiescent SXTs}

The observation of X-ray emission in quiescent SXTs at low
luminosities poses an insurmountable obstacle to models in which the
accretion disk is of the Shakura--Sunyaev type with $r_{\rm in}=3$. On
the one hand, the requirement of Eq. (\ref{e2.5})) can be reconciled
with the deduced mass accretion rate of $\dot M_{\rm
in}\sim10^{11}-10^{12} ~{\rm g\,s^{-1}}$ only if we assume a very low
viscosity in the inner disk, e.g. $\alpha \sim 10^{-12}$ which is
almost equivalent to having no viscosity at all. 
Here we have assumed  $\alpha= const.$. According to the DIM
(see e.g. Cannizzo et al. 1995) $\alpha$ should be of the form $\alpha_0 
\left(H/R\right)^n$, where $H$ is the disk semi-thickness. Cannizzo
et al. (1995) find $\alpha_0=50$, $n=1.5$ and the 
values of  $\alpha$ they obtain are well in excess of those required by  
Eq. (\ref{e2.5}) confirming the failure of the DIM to account for
the observed properties of quiescent SXTs.
On the other hand,
the X--rays have to be produced by a process connected to viscosity,
and if we assume a standard thin disk with $\dot M_{\rm in}\sim
10^{11} ~{\rm g\,s^{-1}}$ as implied by the X-ray observations, the
effective temperature will be so low that the disk will radiate hardly
any X-rays at all.  In other words, the required value of $\alpha$ in
the dwarf--nova type model is unreasonable, and even if we ignore this
problem, it is impossible to fit both the low luminosity and the
temperature of the radiation simultaneously.  The advection-dominated
model of Narayan, McClinotck \& Yi (1996) (hereafter NMY) presents a
solution to this problem.

The fundamental property of advection-dominated accretion flows
(ADAFs) is that the efficiency with which energy is radiated is very
low, so that most of the heat released by viscous friction (or other
processes) is advected into the central black hole and only an almost
infinitesimal amount is radiated away.  Models of ADAFs were first
constructed for optically thick flows with angular momentum (Begelman
1979; Begelman \& Meier 1982; Abramowicz et al. 1986, 1988).
The models
correspond to very high (super-Eddington) accretion rates and have not
yet found an application to real astrophysical objects.

As pointed out by Narayan and Popham (1993), advection may play an
important role also in the case of optically thin flows which are
notoriously non-efficient radiators.  In this case, the viscously
heated gas flows into the black hole on a shorter timescale than its
cooling time and therefore a large fraction of the energy is advected
rather than radiated.  Models of optically thin advection-dominated
flows were constructed in a series of papers by Narayan \& Yi (1994,
1995a,b) and Abramowicz et al (1995), Chen (1995) and Chen et
al. (1995).

The Narayan \& Yi (1995b) version of the ADAF model deals with a
two-temperature plasma which is radiatively cooled by bremsstrahlung,
synchrotron emission and Comptonization. This model was first
successfully applied to the Galactic Center source Sgr A$^{*}$
(Narayan, Yi \& Mahadevan  1995), and soon after 
applied to quiescent SXTs (NMY).
Recently the Narayan \& Yi (1995b) model was applied to the dwarf
active galactic nucleus NGC 4258 (Lasota et al. 1996).

The accretion flow in the NMY model consists of two parts: outside
some radius $\rtrans$ the accreting gas takes the form of a standard,
cold, geometrically thin, Keplerian disk.  However, inside $\rtrans$,
the accretion proceeds in the form of an ADAF.  Here, the gas becomes
extremely hot, with the ions reaching temperatures $\sim 10^{12}$ K
and the electrons going up to $T_e\sim10^{9.5}-10^{10}$ K.  The very
high ion temperature causes the gas to be very thick in the vertical
direction; indeed, the flow is essentially quasi-spherical rather than
disk-like, though it is still partially supported by rotation.  For
simplicity, NMY assumed that the accretion rate is constant throughout
the whole flow.

In the NMY model, the optical and UV radiation is produced by the
outer thin disk, while the inner advection-dominated flow, because of
its very high temperature, produces all the hard radiation in X-rays
and soft $\gamma$-rays.  Even though the same $\dot M$ flows through
the outer and inner zones, the model quite naturally explains the low
luminosity of the X-rays as a consequence of the extremely low
radiative efficiency of the ADAF.  The viscous energy dissipation in
the ADAF is significantly larger than that in the outer thin disk
since most of the gravitational potential energy is released in the
ADAF.  However, virtually all of this energy is advected into the
black hole and only a very tiny fraction is radiated.  This effect is
so strong that the luminosity of the ADAF is actually less than that
of the outer disk.

NMY used their model to fit the spectra of three quiescent BSXTs:
A6020-00, V 404 Cyg and X-ray Nova Muscae 1991.  The fits to the
observed optical--UV to X-ray spectra of the three SXTs are very 
good. The models
depend basically on two parameters, $\dot M /\alpha$ (where $\alpha$
refers to the viscosity parameter in the ADAF and is distinct from
$\alpha_t$ of the outer thin disk) and $\rtrans$.  The other
parameters are either given directly by observations (for example the
black hole mass) or the fits turn out to be insensitive to their
values (for example the ratio of the gas to the total pressure).

\section{The outburst mechanism}

The presence of an inner hot advection-dominated flow removes the
contradiction between the observed X-ray emission and the properties
of a Shakura-Sunyaev type disk.  There remains the question of why
BSXTs go into outbursts.

Could the mechanism be related to the hot inner flow?  Abramowicz et
al (1995) and Narayan \& Yi (1995b) showed that there is a maximum
accretion rate above which there are no optically thin
advection-dominated solutions. In the Narayan \& Yi model the critical
accretion rate is  $\dot M \sim 0.3\alpha^2\dot M_{\rm Edd}$ for
$R<10^3R_{\rm S}$ and $\dot M \sim 0.3\alpha^2\dot M_{\rm Edd}
(R/10^3R_S)^{-1/2}$ for $R>10^3R_{\rm S}$.

\begin{figure}
\epsfxsize=5cm
\epsfbox[18 144 592 718]{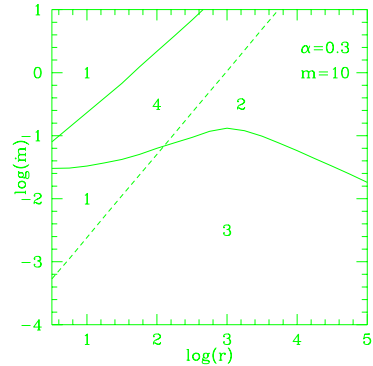}
\caption[]{The regions in the radius (r) $-$ mass accretion
rate (mdot) plane in which the equilibrium flows exist. The upper panel
corresponds to $\alpha=0.3$ and $m=10$. The thin
Shakura-Sunyaev
disk solution is allowed below the dashed line and the optically thick
advection-dominated solutions exist above the upper solid line.
The optically
thin advection dominated flows exist below the lower solid line. For details
see the text and Chen et al. (1995).}
\end{figure}

Figure~2, taken from Narayan (1995) and based on Chen et al. (1995),
represents the structure of all possible accretion solutions for the
case of a 10 $\msun$ black hole.  Label 1 refers to a region where the
only allowed solution is an advection-dominated one, label 2 refers to
a radiatively-cooled zone, and label 3 corresponds to a region where
both types of solutions are allowed.  In region 4 there are no stable
solutions at all.  One could speculate that if the accretion rate at
which matter is brought to the hot ADAF corresponded to region 4, the
system would be forced into a thermal runaway and a limit cycle (see
e.g. Chen et al. 1995).  There is, however, a problem with this
scenario.  NMY found that they needed $\alpha\sim0.1-0.3$ in the
advection-dominated inner zone of their models.  For such $\alpha$,
zone 4 exists only for mass accretion rates above $\sim10^{-2}\dot
M_{\rm Edd}$.  But the mass transfer rates in quiescent SXTs estimated
by NMY correspond to $\dot M \approx (10^{-3} - 10^{-4})\dot M_{\rm
Edd}$.  For such low accretion rates an ADAF solution is always
present.  Furthermore, ADAFs are thermally stable by construction
(Abramowicz et al. 1995, Narayan \& Yi 1995b, Kato et al. 1995) 
The hot inner flows in quiescent SXTs are therefore stable and we
cannot expect to find there the cause of outbursts.

Thus the outburst mechanism in SXTs must be associated with the outer
cold disk which could be, or become thermally unstable.
Below we will investigate the mechanisms that could cause  an
outburst in the outer thin disk.

\subsection{Disk instability triggered by enhanced mass transfer}

In this and the next sections we will study the stability properties
of the outer `standard' thin disk, which extends from the transition
radius $r=\rtrans$ to the outer radius $r=r_{\rm out}$.  As explained
in Section 2, for the disk to be globally unstable, the mass transfer
rate $\dot M_T$ must satisfy at some radius the inequality
\begin{equation}
\dot M_{\rm max}<\dot M_T<\dot M_{\rm min}
\label{e4.1}
\end{equation}
where $\dot M_{\rm max}$ and $\dot M_{\rm min}$ are defined in Section
2.  The range of unstable mass transfer
rates depends on the size of the disk. As an example Figure~3 shows
surface density profiles of equilibrium accretion disks around a $10
\msun$ black hole, where the disk has been truncated at $r=\rtrans$ in
accordance with the NMY model.  The accretion rate $\dot M$ is
constant for each curve. The unstable part of the curve is represented  
by a dotted line, the cold stable configurations correspond to the
solid line.
Models by Hameury et al. (1995) were used in
this calculation with a thin disk viscosity parameter $\alpha_t=0.01$.
Note that this is smaller than the value $\alpha\sim0.1-0.3$ which NMY
found was necessary for a consistent model of the advection-dominated
region.  There is no contradiction in this since there is no reason
for $\alpha$ to be the same in the advection--dominated and standard
regions; one should rather expect $\alpha$ to be lower in the
outer thin disk compared to the inner hot flow.

\begin{figure}
\epsfxsize=8cm
\epsfbox[12 12 576 520]{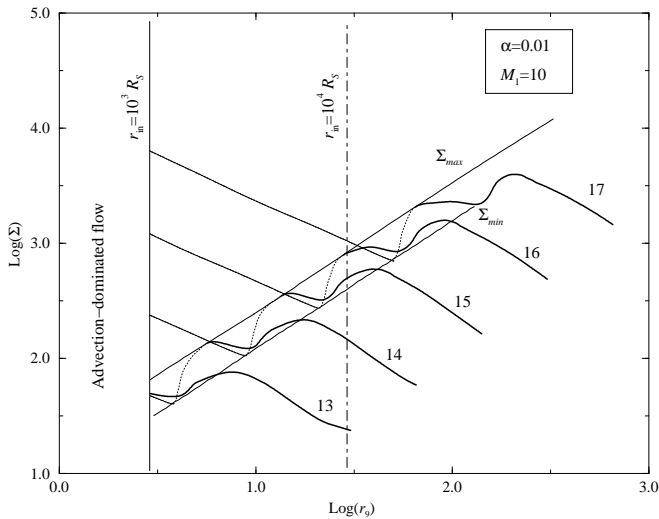}
\caption[]{The surface density profiles of equilibrium accretion
disks around a black hole of $10M_{\odot}$ with $\alpha=0.01$. 
Cold, stable equilibria are represented by solid lines, the unstable
equilibria correspond to the dotted lines.
Inner regions
of thin disks could be truncated at radii $\sim 10^{3-4}$ Schwarzschild
radii ($R_s$) (NMY).}
\end{figure}
A globally stable, cold accretion disk cannot be represented
by a curve that crosses the $\Sigma =\Sigma_{\rm max}$ line.  Assuming
$\Sigma<\Sigma_{\rm max}$ at $r=r_{\rm out}$, we can have a situation
where with decreasing $r$, we reach the condition $\Sigma=\Sigma_{\rm
max}$ at some radius $r=r_{\rm crit}$.  The disk can be in cold
equilibrium only for $r_{\rm crit} < r < r_{\rm out}$.  Continuing to
$r<r_{\rm crit}$, the segment of the curve between $\Sigma_{\rm max}$
and $\Sigma_{\rm min}$ is unstable while the curve to the left of
$\Sigma_{\rm min}$ represents hot equilibria.

For each mass transfer rate $\dot M_T$, there exists therefore a
critical radius $r_{\rm crit}$ outside of which the disk is globally
stable in the cold state.  From Figure~3 we can see that the global
state of the disk will depend on the exact value of $\rtrans$ relative
to $r_{\rm crit}$.  For example, for $\dot M= 10^{15} \gs$, if the
outer disk is truncated at $\rtrans = 10^4$ we have a globally stable
disk whereas if $\rtrans = 10^3$ the disk is globally unstable.  For
the moment we consider $\rtrans$ to be a free parameter that is fixed
by best `fit' to the spectrum of the quiescent SXT in combination with
other parameters such as the black hole mass, $\alpha$ etc. (see NMY).
Physically, of course, the value of $\rtrans$ should be determined by
the process that leads to the formation of the ADAF, such as coronal
evaporation (see e.g. Meyer \& Meyer--Hofmeister 1994) or heating by
diffusive energy transport (Honma 1995).  The theories of these
processes have not yet developed to the point where they can make
robust predictions of $\rtrans$.

Lasota \al (1995) studied a similar problem in the context of dwarf
nova outbursts. Some SU UMa--type dwarf novae show only very rare and
very long superoutbursts and no (or almost no) `normal' outbursts. The
best known system in this class is WZ Sge. Such DN systems are the
closest relatives of BSXTs.  The DIM applied to dwarf novae requires
some modifications before it can be applied to the whole class of
observed events. In particular, the so--called UV--delay problem and
the X--ray and UV emissions from quiescent DNs require a `hole' in the
inner disk regions (see e.g. Meyer \& Meyer--Hofmeister 1994).  This
`hole' can be produced by magnetic disruption if the white dwarf is
(weakly) magnetized (Livio \& Pringle 1992) or it can be the result of
evaporation (Meyer \& Meyer--Hofmeister 1994).

Lasota \al (1995) show that if the inner regions of accretion disks in
quiescent dwarf nova systems are removed, the remaining disk is
globally stable for mass transfer rates $\lta 10^{15}$ g s$^{-1}$.
This implies that (super)outbursts in such systems have to be
triggered by an enhanced mass transfer from the companion. They
suggest that the lack of normal outbursts in WZ Sge results because of
its low mass transfer rate: there are no outbursts because the disk is
stable.  A superoutburst would be triggered by an EMT which would put
the disk into a globally unstable state; in other words,
superoutbursts would be due to a disk instability generated by an
increased mass transfer.  Observations show that the mass transfer
increases prior to and during the superoutburst so that such a hybrid
mechanism could be at work in SU UMa's (Smak 1995, 1996). The
alternative model (see Osaki 1996 for a review) in which
superoutbursts are due to a pure disk instablity, the so--called
tidal--thermal instability, is yet  to be confirmed by 
the observational properties of SU UMa (Smak 1991) and requires 
the viscosity to be extremely low in the case of WZ Sge.

If $\rtrans$ in (some) BSXTs is such that in quiescence the outer thin
disk is globally stable the same reasoning would apply also to these
systems and (some) SXT outbursts could be due to an EMT triggered
disk instability.  

\subsection{Thermal instability in the outer thin disk}

If, for a given mass transfer rate, the inner radius of the outer thin
accretion disk $\rtrans$ is smaller than the critical radius $r_{\rm
crit}$, the disk will be globally unstable and will undergo DN type
outbursts.  We show in this section that we obtain reasonable
recurrence times without any unusual assumption about the value of the
viscosity parameter $\alpha_t$.

In order to estimate properties of the outbursts we closely follow the
steps used by Smak (1993) in the analysis of WZ Sge.

As discussed in Section 2., the disk outburst begins when the density
exceeds somewhere the critical value $\Sigma_{\rm max}$.  The type of
resulting outburst depends on where in the disk this happens.  For a
constant $\dot M_T$, the type of outburst depends on two
characteristic times: the time $t_{\rm accum}$ it takes the matter
accumulating at the outer disk to build up a surface density larger
than $\Sigma_{\rm max}$, and the viscous time $t_{\rm vis}$ which
measures the time it takes matter to diffuse inward and cross the
$\Sigma_{\rm max}$ barrier somewhere nearer to the inner disk
boundary. If $t_{\rm vis} > t_{\rm accum}$ one obtains a so-called
``outside-in" (`type A') outburst beginning at the outer disk edge; in
the opposite case one has an ``inside-out" ('type B') outburst (Smak
1984).

Below, we will consider ``inside-out" outbursts. As discussed by Smak
(1993) these outbursts have longer recurrence times (longer length of
the outburst cycle) than the ``outside in'' outbursts and give also
upper limits on $\alpha_t$. One should note here that although there
are cases where it is claimed that BSXT outbursts began in the inner
regions (Chen et al. 1993), the observational evidence is
far from being conclusive. One should note however that the present
discussion concerns the {\sl outer} thin disk so that even an
``inside-out" outburst in the outer disk would most probably be an
``outside-in" outburst from the point of view of the inner, hot
accretion flow.

Rescaling Eq. (\ref{e2.5}) to the characteristic parameters of the
outer disk we get
\begin{equation}
\dot M \lta 2.3\times10^{14}
\left({\alpha_t\over0.01}\right)^{-0.79}
m_{10}^{2.74}
\left({\rtrans \over10^4}\right)^{3.11}
t_9^{-1} \gs.
\label{e4.3}
\end{equation}
As before, $\alpha_t$ denotes the viscosity parameter in the thin
disk.  If we replace $\dot M$ by the luminosity of the thin disk:
\begin{equation}
L_{\rm disk}\approx{GM\dot M\over R},
\label{e4.4}
\end{equation}
the recurrence time of the instability can be written as
\begin{eqnarray}
\left({t_{rec}\over10^9s}\right) & \lta &    
\left({\alpha_t\over0.01}\right)^{-0.79}
\nonumber \\
& & \left({L_{\rm disk}\over 10^{31}\ergs}\right )^{-1}
\left({m\over10}\right)^{2.74}
\left({\rtrans \over10^4}\right)^{2.11}.  
   \label{e4.5} 
\end{eqnarray}
Eq. (\ref{e4.5}) reveals the severe problem faced by the dwarf-nova type disk
instability model of SXTs and how the problem is resolved in the NMY
two-zone model.  If we set the inner edge of the thin disk to be
$\rtrans=3$, corresponding to the last stable orbit, then for the
observed X-ray luminosity of $\sim10^{31}~{\rm ergs\,s^{-1}}$,
Eq. (\ref{e4.5}) shows that we need an absurdly small
$\alpha_t\sim10^{-12}$.  On the other hand, if we set $\rtrans\sim{\rm
few}\times10^3-10^4$ as found by NMY from their spectral fits, then we
obtain the right recurrence timescale with
$\alpha_t\sim10^{-3}-10^{-2}$,   
consistent with the values of $\alpha_t$ required by the `standard' 
DIM for dwarf novae (Livio \&Spruit 1991). However, our estimated 
value of $\alpha_t$ is very sensitive to the values of $\rtrans$ and $m$, 
so that the values $L_{\rm disk}$, $\rtrans$ and $m$ cannot be much different
from those assumed in Eq. (\ref{e4.5}) for $\alpha_t$ to be not
much smaller than $10^{-3}$. In the case of A0620-00 the optical
luminosity is $\sim 5 \times 10^{32} \ergs$ (McClintock et al. 1995)
so that if one wished to keep $\alpha_t \sim 0.01$,  Eq. (\ref{e4.5}) 
requires $\rtrans > 10^4$. At $r=10^4$  
this optical luminosity corresponds to an accretion
rate $\sim 10^{16} \gs $. Fig.~3 shows however that for  $\rtrans > 10^4$
the outer disk is globally stable so that  Eq.~(\ref{e4.5}) does
not apply because to go into outburst the system requires an EMT.
On the other hand, if we allowed  the value of $\alpha_t$ to 
be as low as $5 \times 10^{-5}$,
which is required for the pure DIM to work in the case of WZ Sge
(Smak 1993), the transition radius could be $\rtrans \sim 10^3$
depending on the value of the black hole mass. 
 
We thus conclude that BSXT outbursts can be triggered by the `normal'
dwarf nova type instability in the outer cold, Keplerian disk.  The
critical factor which allows us to obtain good agreement on the
recurrence time with a ``reasonable'' value of $\alpha_t$ is the
truncation of the thin disk at a large radius $\rtrans$.  
Since such values of $\rtrans > 10^4$ are also required by the 
EMT model, this could be viewed as an independent confirmation 
of the truncated disk model of NMY.

\section{Discussion}

\subsection{Spectral evolution during outburst}

Another feature of the NMY model is that it provides a natural
explanation for the spectral evolution seen in SXTs during outburst
(Narayan 1996).  When the outer disk goes into outburst and the mass
accretion rate increases suddenly, the first response of the system
will be for the $\dot M$ in the inner advection-dominated zone also to
increase proportionately.  Since the luminosity of the inner zone
varies roughly as $\dot M^2$, the X-ray luminosity will go up
enormously.  At the same time, since the electron temperature is very
high, the spectrum will remain hard and there will be emission out to
a few $\times100$ keV.  Model calculations indicate that the photon
spectral index is expected to be $\sim1.5-2$ (Narayan 1996) when the
luminosity reaches $\sim10^{-2}L_{\rm Edd}$.  Meanwhile, the increased
$\dot M$ will probably cause the outer thin disk to diffuse inward on
a viscous time.  Therefore, after some delay, the outer thin disk will
move into the advection-dominated zone, perhaps extending all the way
into the black hole.  When this happens, we will have a very luminous
soft X-ray source with a spectrum consistent with that of a
Shakura-Sunyaev disk.  A0620-00 and Nova Muscae 1991 appear to have
made this transition, whereas V404 Cyg remained a hard source
throughout the outburst.  Perhaps the thin disk did not penetrate all
the way to the center in the latter case.  As the outburst dies down,
we expect these stages to occur in reverse.  First, the thin disk will
shrink back, leaving behind a strong hard X-ray source.  Later, as
$\dot M$ reduces, the system will revert to the quiescent state where
there will be a very low X-ray luminosity and most of the viscous
energy will be advected into the black hole.  The stages described
above, which appear reasonable under the advection-dominated model,
match quite well with observations of several SXTs in outburst.

\subsection{Difficulties}

The NMY model gives a satisfactory description of the quiescent state
of BSXTs, and as we have shown in this paper also provides a
reasonable explanation of the recurrence times of the outbursts and
the spectral evolution during outburst.  All of these were difficult
to explain with the previous  model where the thin disk was
assumed to extend all the way down to the black hole.

There is, however, an inconsistency in the NMY model.  The spectral
`fits' are made under the assumption that the outer disk is
stationary, which means in practice, that a temperature profile $\Teff
\sim r^{-3/4}$ is used to calculate the spectrum.  For the parameters
required for an acceptable spectral fit, the assumed values of $\dot
M$, $\alpha$ and $\rtrans$ are such that $\rtrans<r_{\rm crit}$ so
that a stationary disk is globally unstable.  Such a disk would
therefore be subject to the dwarf-nova type instability, which means
that in the quiescent state the accretion rate would not be constant
with radius.  This implies a self-inconsistency in the NMY model.

A further problem arises from the $T_{\rm eff}$ required to fit the
optical spectra.  Taking for the effective temperature (cf. Frank et
al. 1992)
\begin{equation}
\Teff(R)= \left[{3GM\dot M\over 8\pi \sigma R^3}\right]^{1/4},
\label{e5.1}
\end{equation}
and replacing $\dot M$ by the expression given in Eq. (\ref{e4.3}) we
see that
\begin{equation}
\Teff(r)\approx 2230 \left({\alpha_t\over 0.01}\right)^{-0.2} m_{10}^{0.19}
\left({\rtrans \over 10^4}\right)^{0.03} t_9^{-1/4} \ K,
\label{e5.2}
\end{equation}
which shows that the effective temperature in a quiescent dwarf nova
disk is practically independent of the radius.  
For $m=1$ and a recurrence time $\sim$ 60 days, typical parameters of
a dwarf nova, Eq. (\ref{e5.2}) gives (for $\alpha_t = 0.01$) $\Teff
\sim 5400\ K$, very close to the observed values (Wood et al. 1986).

The problem arises when we apply the same formula to quiescent SXTs.
For SXT parameters, the effective temperature turns out to be $\sim
2000$ K which is too low to agree with observations.  Both in A0620-00
and V404 Cyg, the optical and UV data imply significantly higher
temperatures.  It should be emphasized that this problem is not
peculiar to the NMY model, but will be faced by any SXT model which
invokes a dwarf nova type non-steady quiescent accretion disk.

A possible explanation for the higher disk temperature could be that
the outer disk is optically thin.  The strong H$\alpha$ line in the
spectra of these disks suggests that a substantial part of the
emission comes from optically thin gas.  It is then natural for the
color temperature to be higher than the effective temperature.

Finally, one should note that the failure of the DIM to account for
the properties of quiescent SXTs has been also noticed  
by Cannizzo et al. (1995) and Kim et al. (1996) who suggested
`non-standard' inner disk structures
that do not invoke advection.

\section{Conclusions}

The detailed mechanism by which BSXTs go into outburst will depend on
the values of $\dot M$, $\alpha$ and $r_{\rm tr}$. These parameters,
or rather the ratio $\dot M /\alpha$ and $r_{\rm tr}$, are the free
parameters which NMY used to fit their models to observed spectra of
quiescent SXTs. The physical conditions in the outer disk are very
similar to those that are encountered in accretion disks of SU UMa systems.
For this reason also the problems with the origin of the outbursts
(or rather superoutbursts) are the same.

At the low accretion rates that are required by observations of BSXTs
in quiescensce, the accretion disk may be globally stable for
$r \gta \rtrans$. In this case only an increase in the mass
transfer from the secondary can bring the disk into an unstable state
and trigger the outburst. To make the model complete one should find why
the secondary increases its transfer rate on timescales typical
of SXTs. One should note however that variations of the transfer rate
on various timescales {\sl are observed} in many close binary systems.

When $\dot M$, $\alpha$ and $r_{\rm tr}$ correspond to a
globally unstable outer disk configuration, a DN type instability will
necessarily occur.  The effect of the outburst on the inner hot ADAF
should be similar to that of a EMT triggered outburst since both types
of outbursts are basically due to the same thermal instability.

It is not impossible that both types of mechanisms are at work in BSXTs,
in different systems and/or in the same systems but at different epochs.

The main result of this paper is that the disk instability model works
much better for SXTs if the thin disk is truncated at a large
transition radius $\rtrans$ as in the NMY model rather than having the
disk extending down to $r=3$ as in the  model by Mineshige \&
Wheeler (1989).  The region inside the radius of truncation would then
be filled with a very hot, optically thin, ultra-low radiative
efficiency, advection-dominated flow.  NMY showed that such a model
gives a good fit to the spectra of quiescent SXTs, while Narayan
(1996) suggested that it may be consistent with the spectral evolution
observed in SXTs during outburst.  In this paper, we show that the
recurrence times of SXT outbursts are satisfactorily predicted with a
reasonable value of the viscosity parameter
$\alpha_t\sim10^{-3}-10^{-2}$ in the outer disk (or $\sim 10^{-5}$
in the worst case).
In contrast, in the
dwarf--nova type model, one requires an implausibly low value
$\alpha_t\sim10^{-12}$.  
 
One interesting point to bear in mind is that it is possible to have
binaries where $\dot M$, $\alpha$ and $\rtrans$ correspond to a
globally stable outer disk in a cold low state, and where in addition
the secondary star for whatever reason does not have any tendency to
exhibit a large amplitude EMT instability.  In such a system, we would
have a black hole binary which is permanently in the quiescent state,
with most of the accretion energy being swallowed by the black hole.
The system would be extremely dim in X-rays and quite under-luminous
even in the optical and would be very hard to detect.  In principle,
the Galaxy could have large numbers of such nearly silent black holes,
which would be the binary equivalent of the underluminous galactic
nuclei discussed by Fabian \& Rees (1995).

We are grateful to Jean-Marc Hur\'e for providing us with
Figures 1 and 3. We thank Jean-Marie Hameury for discussion and comments
and Craig Wheeler for helpful critical remarks.
RN was supported in part by NASA grant NAG 5-2837 (to the Center
for Astrophysics) and NSF grant PHY 9407194 (to the Institute
for Theoretical Physics). IY acknowledges support from SUAM Foundation.

\end{document}